\documentclass{article}

\usepackage[preprint]{neurips_2020_ml4ps}

\usepackage[utf8]{inputenc} 
\usepackage[T1]{fontenc}    
\usepackage{hyperref}       
\usepackage{url}            
\usepackage{booktabs}       
\usepackage{amsfonts}       
\usepackage{nicefrac}       
\usepackage{microtype}      
\usepackage{multirow}
\usepackage{ctable}
\usepackage{multicol}

\title{Domain adaptation techniques for improved cross-domain study of galaxy mergers}

\author{
\hspace*{-2mm}
A. \'Ciprijanovi\'c\\
\hspace*{-2mm}
\texttt{aleksand@fnal.gov}\\
\hspace*{-2mm}
Fermi National Accelerator Laboratory\And

\hspace*{4mm}
D. Kafkes\\
\hspace*{4mm}
\texttt{dkafkes@fnal.gov}\\
\hspace*{4mm}
Fermi National Accelerator Laboratory\And

\hspace*{-2mm}
S. Jenkins\\
\hspace*{-2mm}
\texttt{sydneyjenkins@uchicago.edu}\\
\hspace*{-2mm}
Department of Astronomy and Astrophysics,\\University of Chicago\And

\hspace*{-1mm}
K. Downey\\
\hspace*{-1mm}
\texttt{kdowney@uchicago.edu}\\
\hspace*{-1mm}
Department of Astronomy and Astrophysics,\\ 
\hspace*{-1mm}
University of Chicago\And

\hspace*{-7mm}
G. N. Perdue\\
\hspace*{-7mm}
\texttt{perdue@fnal.gov}\\
\hspace*{-7mm}
Fermi National Accelerator Laboratory\And

\hspace*{2mm}
S. Madireddy\\
\hspace*{2mm}
\texttt{smadireddy@anl.gov}\\
\hspace*{2mm}
Argonne National Laboratory\And

\hspace*{9mm}
T. Johnston\\
\hspace*{9mm}
\texttt{johnstonjt@ornl.gov}\\
\hspace*{9mm}
Oak Ridge National Laboratory\And

\hspace*{9mm}
B. Nord\\
\hspace*{9mm}
\texttt{nord@fnal.gov}\\
\hspace*{9mm}
Fermi National Accelerator Laboratory,\\
\hspace*{9mm}
Kavli Institute for Cosmological Physics,\\ 
\hspace*{9mm}
University of Chicago \\
\hspace*{9mm}
Department of Astronomy and Astrophysics,\\
\hspace*{9mm}
University of Chicago \\
}

\begin{document}

\maketitle

\begin{abstract}
In astronomy, neural networks are often trained on simulated data with the prospect of being applied to real observations. Unfortunately, simply training a deep neural network on images from one domain does not guarantee satisfactory performance on new images from a different domain. The ability to share cross-domain knowledge is the main advantage of modern deep domain adaptation techniques. Here we demonstrate the use of two techniques --- Maximum Mean Discrepancy (MMD) and adversarial training with Domain Adversarial Neural Networks (DANN) --- for the classification of distant galaxy mergers from the Illustris-1 simulation, where the two domains presented differ only due to inclusion of observational noise. We show how the addition of either MMD or adversarial training greatly improves the performance of the classifier on the target domain when compared to conventional machine learning algorithms, thereby demonstrating great promise for their use in astronomy.

\end{abstract}

\section{Introduction}
\label{sec:intro}

The study of galaxy mergers plays an important role in discerning the existence of different types of galaxies, documenting their origins, and furthering our understanding of the evolution of the entire universe and its appearance today. It has been shown that machine learning can greatly advance the study of merging galaxies [1,16,12,4], yet without the ability to connect the knowledge obtained from disparate large-scale simulations and astronomical surveys, we are at a significant disadvantage towards the goal of harnessing all available data.

Images produced by astronomical simulations are made to mimic real observations from a particular telescope, but even the slightest differences (which are unavoidable) can cause a classification model trained on simulated images to perform substantially worse on related real data. One such example can be found in [12], where authors used a data set of merging galaxies from the EAGLE simulation [13] made to mimic SDSS observations and real SDSS images [20]. In this paper, authors show that the performance of the classifier on the task of distinguishing merging from non-merging galaxies trained on one data set had much lower accuracy when classifying the other data set: $53-65\%$, with the classifier trained on real SDSS images and then applied to EAGLE simulation images yielding particularly poor performance. Similarly, in [4] authors work with distant merging galaxies from the Illustris-1 cosmological simulation [17]. Here it was shown that, even in the case where the domains only differ due to inclusion of noise to mimic the Hubble Space Telescope observations, the accuracy of classification in the domain the model was not trained on hovers around $50\%$, no better than random guessing. These two examples are indicative of the need for domain adaptation techniques to be applied in astrophysical contexts.

Domain adaptation techniques are used to detect the shift between source domain and target domain distributions [5,8,19]. This functionality is very useful in situations often found in astronomy where models are trained on labeled simulations and then applied to unlabeled real data. In this paper we apply two domain adaptation techniques as transfer loss: MMD [15,7] and adversarial training using DANNs [6]. We demonstrate both techniques on a data set similar to the one from [4]: we use simulated distant merging galaxies from Illustris-1 at redshift $z=2$, both without (source) and with (target) the addition of random sky shot noise to mimic observations from the Hubble Space Telescope. We also test two networks for a comparison of results across architectures: DeepMerge, a simple network for classification of galaxies presented in [4], as well as the more complex and well-known ResNet18 [9]. In both cases, we show that the use of the identified domain adaptation techniques lead to a significant improvement in the performance of the classifier on the target domain.

\section{Methods}
\label{sec:methods}

In our experiments, the neural network is trained using the total loss ${\cal L}_\mathrm{Total}$:
\begin{equation}
{\cal L}_\mathrm{Total} = {\cal L}_\mathrm{CL} + \lambda_\mathrm{TL}{\cal L}_\mathrm{TL},\label{eq:total}
\end{equation}
where we label ${\cal L}_\mathrm{CL}$ and ${\cal L}_\mathrm{TL}$ as classifier loss (cross-entropy) and transfer loss. The effects of MMD and adversarial training are applied through the latter which is weighted by constant $\lambda_\mathrm{TL}$.

In order to detail MMD and adversarial training below, we first introduce the following conventions: we denote the source and target domains as ${\cal D}_\mathrm{s}$ and ${\cal D}_\mathrm{t}$ and their respective distributions as ${\cal P}_\mathrm{s}$ and ${\cal P}_\mathrm{t}$. Since the source domain data are labeled, we have $n_\mathrm{s}$ pairs of images and labels $\{\mathbf{x}_\mathrm{s}, \mathbf{y}_\mathrm{s}\}$, while in the unlabeled target domain we only have $n_\mathrm{t}$ unlabeled images $\mathbf{x}_\mathrm{t}$. Finally, data from both domains are associated with domain labels: $d_\mathrm{s}$ for source and $d_\mathrm{t}$ for target domain.

\paragraph{Maximum Mean Discrepancy (MMD)} as a transfer loss works by minimizing the distance between the means of ${\cal P}_\mathrm{s}$ and ${\cal P}_\mathrm{t}$. While it is possible to estimate ${\cal P}_\mathrm{s}$ and ${\cal P}_\mathrm{t}$, in practice, no computationally expensive density estimation is necessary [11]. Instead, kernel methods may be applied to determine their means for subtraction and an optimization is undertaken in an RKHS (Reproducing Kernel Hilbert Space):

\begin{equation}
\label{eq:mmd}
    D( {\cal P}_\mathrm{s}, {\cal P}_\mathrm{t}, {\cal F}) :=
    \sup_{||f|| \leq 1}\mathbb{E}_{\cal P_\mathrm{s}}[\langle k(x_\mathrm{s}), f \rangle] - \sup_{||f|| \leq 1}\mathbb{E}_{\cal P_\mathrm{t}}[\langle k(x_\mathrm{t}), f \rangle] = \sup_{f}\langle \mu_\mathrm{s} - \mu_\mathrm{t}, f \rangle,
\end{equation}

\noindent where $D$ denotes the kernel distance as a proxy for mean discrepancy, $x_\mathrm{s}$ and $x_\mathrm{t}$ are random variables drawn from ${\cal P}_\mathrm{s}$ and ${\cal P}_\mathrm{t}$ respectively, $f$ closely resembles a cumulative distribution function, and the simplification follows from the reproducibility property of RKHS: $\langle f, k(x, \cdot) \rangle = f(x)$ [8,11]. While in practice, $k$ can be considered a general kernel, we follow [21] where $k$ is a linear combination of multiple Gaussian Radial Basis Function (RBF) kernels to extend across a range of mean embeddings.

By this definition, if ${\cal P_\mathrm{s}} \neq {\cal P_\mathrm{t}}$, then there must exist some $f$ such that the distance between the two means is maximized [11]. Clearly, the inner product is maximized for the identity $\langle a,a \rangle = 1$. Therefore, $f$ must equal $\mu_\mathrm{s}$ - $\mu_\mathrm{t}$ to maximize the mean discrepancy [11]. This leaves us with the final transfer loss, after some discretization:

\begin{equation}
{\cal L}_\mathrm{TL}=\frac{1}{m(m-1)} \sum\limits_{i!=j} k(x_\mathrm{s}(i), x_\mathrm{s}(j)) - k(x_\mathrm{s}(i), x_\mathrm{t}(j))-
k(x_\mathrm{t}(i), x_\mathrm{s}(j)) + k(x_\mathrm{t}(i), x_\mathrm{t}(j)),
\end{equation}

where $m$ is the total number of samples. Here the distance is expressed as the difference between the self-similarities of source ($\mathbb{E}_{\cal P_\mathrm{s}}[k(x_\mathrm{s},x_\mathrm{s}')]$) and target ($\mathbb{E}_{\cal P_\mathrm{t}}[k(x_\mathrm{t},x_\mathrm{t}')]$) domains and their cross-similarity ($2 \mathbb{E}_{\cal P_\mathrm{s,t}}[k(x_\mathrm{s},x_\mathrm{t})]$).

\paragraph{Domain adversarial training} employs a DANN to distinguish between the source and target domains [6]. DANNs are comprised of three parts: feature extractor ($G$), label predictor ($L$), and domain classifier ($D$). Like all deep learning models applied for the classification of images, the feature extractor uses convolutional layers to extract features from images, while the label predictor has dense layers which output the class label. In contrast, the domain classifier, which is also built from dense layers, is unique to DANNs and is used to predict the domain labels.

The domain classifier is added after the feature extractor as a parallel branch to the label predictor, and it includes a gradient reversal layer which maximizes the loss for this branch of the neural network.This leads to the feature extractor being trained with an adversarial objective to confuse the domain classifier. When the domain classifier fails to discriminate between the domains, domain-invariant features have been found and the classifier can then be successfully applied across the two domains.

Domain classifier loss is calculated as:
\begin{equation}
{\cal L}_\mathrm{d} = \frac{1}{n_\mathrm{s}} \sum\limits_{\mathbf{x}_\mathrm{s}\in{\cal D}_\mathrm{s}} l(D(G(\mathbf{x}_\mathrm{s})),d_\mathrm{s})   + \frac{1}{n_\mathrm{t}}  \sum\limits_{\mathbf{x}_\mathrm{t}\in{\cal D}_\mathrm{t}} l(D(G(\mathbf{x}_\mathrm{t})),d_\mathrm{t}),
\end{equation}
where $l(D(G(\mathbf{x}_\mathrm{s})),d_\mathrm{s})$ and $l(D(G(\mathbf{x}_\mathrm{t})),d_\mathrm{t})$ are the output probabilities for the source domain and target domain labels respectively, calculated using cross-entropy loss. Finally, we designate this domain classifier loss as our transfer loss: ${\cal L}_\mathrm{TL} = \mathrm{max}_D \{ -{\cal L}_\mathrm{d} \}$.

\section{Data}
We use a similar data set as in [4], where authors extract galaxies at redshift $z=2$ from Illustris-1. Our dataset differs only by the addition of one more filter to get three channel images: ACS F814W, NC F356W, WFC3 F160W. The source domain includes images from Illustris-1 convolved with a model point-spread function (PSF), while the target domain additionally includes random sky shot noise. More details about the data set can be found in [4]. The source and target domains contain $8120$ merger and $7306$ non-merger images ($75\times 75$ pixels). We divide these data sets into training, validation, and testing samples: $70\%:10\%:20\%$.

\section{Experiments}
\label{sex:experiments}

We present the performance of both domain adaptation techniques in two neural network architectures: the DeepMerge architecture introduced in [4] and the more complex ResNet18 [9]. Both networks are trained for the task of distinguishing between two classes of objects: merging and non-merging galaxies. We first train our two classifier networks without any domain adaptation on the pristine labelled source data only. We then train with the addition of MMD and domain adversarial training. Both deep domain adaptation techniques involve training with both the pristine labelled source data and the unlabelled noisy target data. Finally, we evaluate all three training configurations on both the source and target domain data.

In all experiments, we use the Adam optimizer [10] with implemented "one-cycle" scheduling, which was shown to lead to much faster convergence of training accuracy [14]. We also include early stopping, to prevent overfitting. Additionally, our choice of hyperparameters was informed from the results of a hyperparameter search using DeepHyper [3,2], with only one of the domain adaptation techniques employed for each network. Furthermore, in all experiments we use the same fixed random seed (1) to shuffle images and initialize network weights in order to ensure result consistency.

\section{Results}
\label{sec:results}

Resulting source and target classification accuracies of merging and non-merging galaxies for the three experiments detailed above are given in Table~\ref{table:performance}. We designate our base for improvement as the case without domain adaptation, where, as expected, test accuracy on source images is high, while the classifier performs much worse on target domain images.

\begin{table*}[htpb]
   \centering
   \noindent\begin{minipage}[b]{\textwidth}
   \centering
    \caption{Classification accuracies of DeepMerge and ResNet18 architectures, on source and target domain test sets, without domain adaptation (first row) and when domain adaptation techniques are used during training (all other rows).}
  \label{table:performance}
  \centering
  \begin{tabular}{l c c c c}
    \multicolumn{2}{c}{}  & \multicolumn{2}{c}{}  \\
\multirow{2}{*}{}      &   \multicolumn{2}{c}{Source}   &   \multicolumn{2}{c}{Target}\\
                                                    & DeepMerge & ResNet18 & DeepMerge & ResNet18   \\\hline
\multirow{1}{*}{No Domain Adaptation}               & $84.8\%$  & $81.1\%$ & $52.0\%$  & $60.5\%$    \\
\multirow{1}{*}{MMD}                               & $86.9\%$ & $90.4\%$ & $76.6\%$  & $\mathbf{73.8}\%$     \\
\multirow{1}{*}{Adversarial}                     & $87.4\%$  & $92.3\%$ & $\mathbf{78.6\%}$  & $71.6\%$           \\ \hline
\end{tabular}
\end{minipage}
\end{table*}

\begin{figure*}[h]
    \includegraphics[width=.5\linewidth]{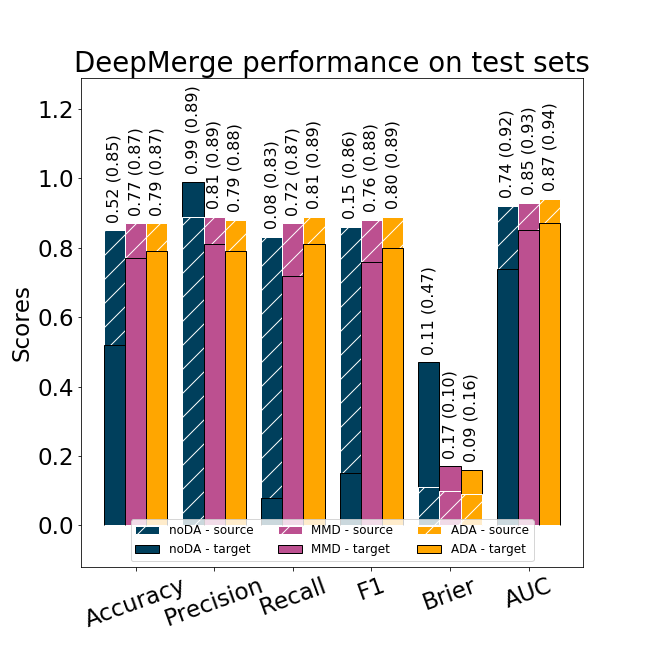}\hfill
    \includegraphics[width=.5\linewidth]{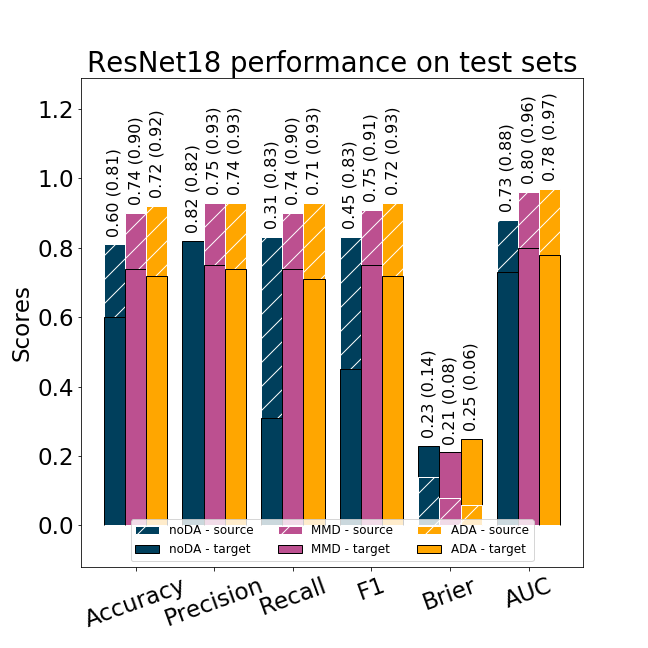}\\ 
    \includegraphics[width=.5\linewidth]{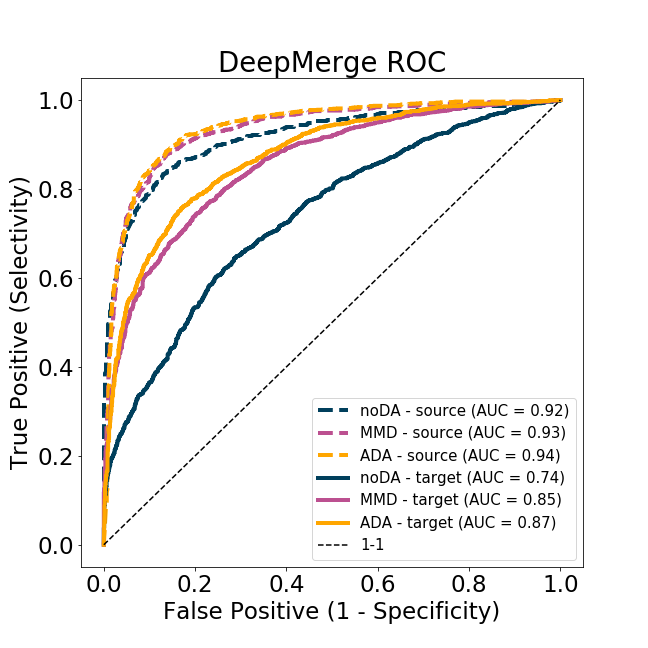}\hfill
	\includegraphics[width=.5\linewidth]{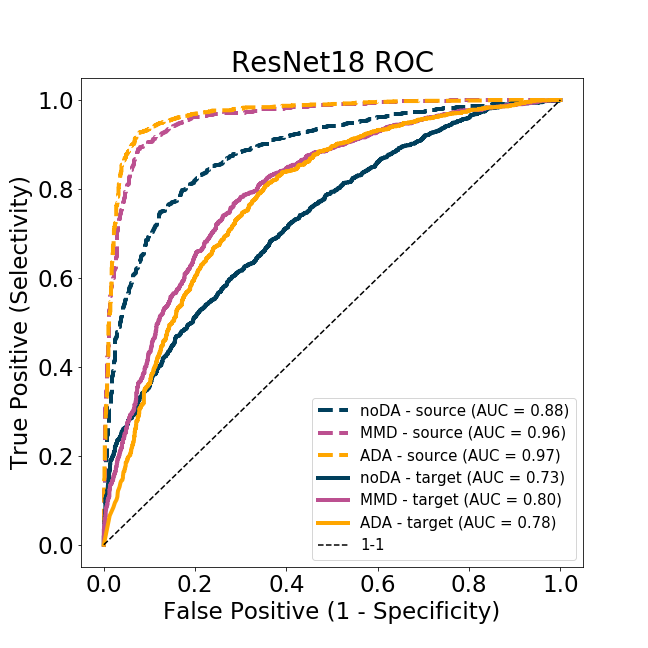}
    \caption{\emph{Top}: Performance metrics, for no domain adaptation experiment (navy), MMD (violet) and adversarial training (orange), on source (dashed bars) and target (solid bars) domain test set. Values above bars are given as: target (source). We plot values for accuracy, precision, recall, F1 score, Brier score and AUC, for both DeepMerge (left) and ResNet18 (right). \emph{Bottom}: ROC curves (AUC values in legend) for classification of source (dashed lines) and target domain (solid lines) test set of images with same color and hatching scheme for both DeepMerge (left) and ResNet18 (right).}
    \label{fig:1}
\end{figure*}

While we expected that with domain adaptation we would see a slight decrease in performance in the source domain in order to compensate for the recognition of shared features across domains, what we actually observe is an increase in performance for the source domain accuracy. We believe this is due to the regularizing effect of the additional transfer loss included in MMD and adversarial training, which assists in preventing overfitting on the source training data set which allows longer training of the model. As expected, the target domain classification accuracy improves in both training with MMD and adversarial training. Additional metrics for performance comparison on the source (dashed bars) and target (solid bars) domain test set of images are presented in the top row of Figure~\ref{fig:1}. Here training without domain adaptation is navy, MMD is violet, and adversarial training is orange. The bottom row of Figure~\ref{fig:1} shows the comparison of ROC curves (Receiver Operating Characteristics) for source and target test set of images, both with and without domain adaptation, with the same color and hatching scheme.

Furthermore, between the two networks, we posit that the smaller improvements made with ResNet18 in the target domain are the result of the much greater architecture complexity --- two orders of magnitude more trainable parameters than DeepMerge --- making it more susceptible to overfitting on the source domain. Early stopping patience and weight decay were invoked to tackle this issue, but resulted in only marginal improvements. Since we found the methods to be extremely sensitive to the hyperparameters chosen, we feel there is still room for further improvement with the choice of optimal parameters (the hyperparameter search was not run on the task without domain adaptation) and perhaps even network pruning.

While the results are quite sensitive to the choice of hyperparameters, we report that they are robust to random seed choice. We ran 10 different random seeds for all experiments with DeepMerge, and did not see significant deviation in performance outside of the target test set without domain adaptation (which is expected since the classifier does not work). We report the following mean $\mu$ and standard deviation $\sigma$ for each experiment: no domain adaptation source ($83.6\pm 0.7\%$) and target ($57.0 \pm 5.2\%$); MMD source ($86.6 \pm 1.3\%$) and target ($77.3 \pm 0.6\%$); adversarial training source ($86.5 \pm 0.9\%$) and target ($78.9 \pm 0.5\%$).

\section{Conclusion}
\label{sec:conclusion}

Astronomy is entering the big data era with a plethora of simulations and many ongoing and future large surveys. Without the ability to connect the knowledge obtained from these different domains, we are at a significant disadvantage to harness all available data. In this paper, we show the promise for the use of domain adaptation techniques, like MMD and adversarial training, in astronomy to substantially improve the performance of a source-trained model on a new and often unlabeled target domain data set. While the scope of this paper is to demonstrate the efficacy of MMD and adversarial training in the case with two simulated domains that differ only due to the inclusion of observational noise, our future work will address results of these techniques applied to simulated and real observational data.

We acknowledge that, while domain adaptation techniques can be very powerful, their ultimate performance depends on the similarity between the source and target domains. To ensure the best possible performance across domains in astronomy--- particularly when training with a simulated source domain and real target domain, the simulated data must be made to mimic the target domain and should contain only in-distribution objects for classification. Differences due to the limitations of the simulator, or differences in the noise and other observational effects added to the simulated images, can then be addressed by domain adaptation. It is for this reason that we firmly believe that studying and refining domain adaptation techniques will prove crucial to successfully deploying deep learning models in astronomy.

\clearpage

\section*{Broader Impact}

This research will impact the astronomy community but also the wider scientific community, since domain transfer problems are very common in many areas of research. In experimental high-energy physics, astronomy and cosmology, biotechnology, etc. research often involves studying physical processes using simulations either before real data becomes available or in conjunction with it. This paper demonstrates the capability of domain adaptation techniques as an important tool in this process.

\begin{ack} 
The authors of this paper have committed themselves to performing this work in an equitable, inclusive, and just environment, and we hold ourselves accountable, believing that the best science is contingent on a good research environment.
We acknowledge the Deep Skies Lab as an open community of multi-domain experts and collaborators. This community was important for the development of this project. 

This manuscript has been supported by Fermi Research Alliance, LLC under Contract No. DE-AC02-07CH11359 with the U.S. Department of Energy, Office of Science, Office of High Energy Physics. This research has been partially supported by the High Velocity Artificial Intelligence grant as part of the Department of Energy High Energy Physics Computational HEP sessions program.

This research used resources of the Argonne Leadership Computing Facility at Argonne National Laboratory, which is a user facility supported by the Office of Science of the U.S. Department of Energy under contract DE-AC02-06CH11357.

\end{ack}

\medskip
\small
\section*{References}

[1] Sandro Ackermann, Kevin Schawinski, Ce Zhang, Anna K. Weigel, and M. Dennis Turp. Using transferlearning to detect galaxy mergers. {\it Monthly Notices of the Royal Astronomical Society}, 479(1):415–425, September 2018.

[2] Prasanna Balaprakash, Romain Egele, Misha Salim, Stefan Wild, Venkatram Vishwanath, Fangfang Xia, Tom Brettin, and Rick Stevens. Scalable reinforcement-learning-based neural architecture search forcancer deep learning research. In {\it Proceedings of the International Conference for High Performance Computing, Networking, Storage and Analysis}, SC ’19, New York, NY, USA, 2019. Association for Computing Machinery.

[3] Prasanna Balaprakash, Salim Misha, Tomas D. Uram, Venkatram Vishwanath, and Stefan M. Wild. Deep-hyper: Asynchronous hyperparameter search for deep neural networks. In {\it 2018 IEEE 25th International Conference on High Performance Computing (HiPC)}, pages 42–51, 2018.

[4] Aleksandra \'Ciprijanovi\'c, Gregory F. Snyder, Brian Nord, and Joshua E. G. Peek. DeepMerge: Classifying high-redshift merging galaxies with deep neural networks. {\it Astronomy and Computing}, 32:100390, July2020.

[5] Gabriela Csurka. A Comprehensive Survey on Domain Adaptation for Visual Applications, pages 1–35. Springer International Publishing, Cham, 2017.

[6] Yaroslav Ganin, Evgeniya Ustinova, Hana Ajakan, Pascal Germain, Hugo Larochelle, Fran{\c{c}}ois Laviolette, Mario March, and Victor Lempitsky. Domain-adversarial training of neural networks. {\it Journal of Machine Learning Research}, 17(59):1–35, 2016.

[7] Arthur Gretton, Karsten Borgwardt, Malte J. Rasch, Bernhard Sch{\"o}lkopf, and Alexander J. Smola.  A Kernel Two-Sample Test. {\it Journal of Machine Learning Research}, 13(25):723–773, 2012.

[8] Arthur Gretton, Dino Sejdinovic, Heiko Strathmann, Sivaraman Balakrishnan, Massimiliano Pontil, Kenji Fukumizu, and Bharath K. Sriperumbudur.  Optimal kernel choice for large-scale two-sample tests.  In F. Pereira, C. J. C. Burges, L. Bottou, and K. Q. Weinberger, editors, {\it Advances in Neural Information Processing Systems} 25, pages 1205–1213. Curran Associates, Inc., 2012.

[9] Kaiming He, Xiangyu Zhang, Shaoqing Ren, and Jian Sun. Deep Residual Learning for Image Recognition. arXiv e-prints, page arXiv:1512.03385, December 2015.
 
[10] Diederik P. Kingma and Jimmy Ba. Adam: A method for stochastic optimization. {\it CoRR}, abs:1412.6980, 2015.

[11] Sinno Jialin Pan, I. W. Tsang, J. T. Kwok, and Qiang Yang. Domain adaptation via transfer component analysis. {\it Trans. Neur. Netw.}, 22(2):199–210, February 2011.

[12] William . J. Pearson, Lingyu Wang, James W. Trayford, Carlo E. Petrillo, and Floris F. S. van der Tak. Identifying galaxy mergers in observations and simulations with deep learning. {\it Astronomy \& Astrophysics}, 626:A49, June 2019.

[13] Joop Schaye, Robert A. Crain, Richard G. Bower, Michelle Furlong, Matthieu Schaller, Tom Theuns,Claudio Dalla Vecchia, Carlos S. Frenk, I. G. McCarthy, John C. Helly, Adrian Jenkins, Y. M. Rosas-Guevara, Simon D. M. White, Maarten Baes, C. M. Booth, Peter Camps, Julio F. Navarro, Yan Qu, Alireza Rahmati, Till Sawala, Peter A. Thomas, and James Trayford. The EAGLE project: simulating the evolution and assembly of galaxies and their environments. {\it Monthly Notices of the Royal Astronomical Society}, 446(1):521–554, January 2015.

[14] Leslie N. Smith and Nicholay Topin.  Super-convergence:  very fast training of neural networks using large learning rates. In Tien Pham, editor, {\it Artificial Intelligence and Machine Learning for Multi-Domain Operations Applications}, volume 11006, pages 369 – 386. International Society for Optics and Photonics,SPIE, 2019.

[15] Alexander J. Smola, Arthur Gretton, Le Song, and Bernhard Sch{\"o}lkopf.   A hilbert space embedding for distributions.   In {\it Algorithmic Learning Theory}, Lecture Notes in Computer Science 4754, pages 13–31, Berlin, Germany,October 2007. Max-Planck-Gesellschaft, Springer.

[16] Gregory F. Snyder, Vicente Rodriguez-Gomez, Jennifer M. Lotz, Paul Torrey, Amanda C. N. Quirk, Lars Hernquist, Mark Vogelsberger, and Peter E. Freeman.  Automated distant galaxy merger classifications from Space Telescope images using the Illustris simulation. {\it Monthly Notices of the Royal Astronomical Society}, 486(3):3702–3720, July 2019.

[17] Mark Vogelsberger, Shy Genel, Volker Springel, Paul Torrey, Debora Sijacki, Dandan Xu, Greg Snyder, Dylan Nelson, and Lars Hernquist. Introducing the Illustris Project: simulating the coevolution of dark and visible matter in the Universe. {\it Monthly Notices of the Royal Astronomical Society}, 444(2):1518–1547, October 2014.

[18] Mei Wang and Weihong Deng.  Deep visual domain adaptation: A survey. {\it Neurocomputing}, 312:135 –153, 2018.

[19] Garrett Wilson and Diane J. Cook. A survey of unsupervised deep domain adaptation. {\it ACM Trans. Intell. Syst. Technol.}, 11(5), July 2020.

[20] Donald G. York, J. Adelman, Jr. Anderson, John E., Scott F. Anderson, James Annis, Neta A. Bahcall, J. A. Bakken, Robert Barkhouser, Steven Bastian, Eileen Berman, William N. Boroski, and et al., SDSS Collaboration. The Sloan Digital Sky Survey: Technical Summary. {\it The Astronomical Journal}, 120(3):1579–1587, September 2000.

[21] Yinghua Zhang, Yu Zhang, Ying Wei, Kun Bai, Yangqiu Song, and Qiang Yang.  Fisher Deep Domain Adaptation. arXiv e-prints, page arXiv:2003.05636, March 2020.

\end{document}